\documentclass[prd,12pt,tightenlines,superscriptaddress,showpacs,%
preprintnumbers,nofootinbib,amsmath]{revtex4}

\usepackage{bm}

\begin{document}

\preprint{CECS-PHY-04/06}
\preprint{hep-th/0403229}

\title{Four Parametric Regular Black Hole Solution}

\author{Eloy Ay\'on--Beato}\email{ayon@cecs.cl}
\affiliation{Centro~de~Estudios~Cient\'{\i}ficos~(CECS),%
~Casilla~1469,~Valdivia,~Chile.}
\affiliation{Departamento~de~F\'{\i}sica,%
~Centro~de~Investigaci\'on~y~de~Estudios~Avanzados~del~IPN,\\
~Apdo.~Postal~14--740,~07000,~M\'exico~D.F.,~M\'exico.}
\author{Alberto Garc\'\i a}\email{aagarcia@fis.cinvestav.mx}
\affiliation{Departamento~de~F\'{\i}sica,%
~Centro~de~Investigaci\'on~y~de~Estudios~Avanzados~del~IPN,\\
~Apdo.~Postal~14--740,~07000,~M\'exico~D.F.,~M\'exico.}

\date{\today}

\begin{abstract}
We present a regular class of \emph{exact} black hole solutions of
Einstein equations coupled with a nonlinear electrodynamics
source. For weak fields the nonlinear electrodynamics becomes the
Maxwell theory, and asymptotically the solutions behave as the
Reissner--Nordstr\"om one. The class is endowed with four
parameters, which can be thought of as the mass $m$, charge $q$,
and a sort of dipole and quadrupole moments $\alpha $ and $\beta$,
respectively. For $\alpha\geq3$, $\beta\geq4$, and
$|q|\leq2s_\mathrm{c}m$ the corresponding solutions are regular
charged black holes. For $\alpha=3$, they also satisfy the weak
energy condition. For $\alpha=\beta=0$ we recover the
Reissner--Nordstr\"om singular solution and for $\alpha=3$,
$\beta=4$ the family includes a previous regular black hole
reported by the authors.
\end{abstract}


\maketitle

One of the active research topics on black hole Physics in recent
years has been related with its interior behavior. It is clear now
that black holes are not necessarily singular and several examples
of regular \emph{exact} black hole solutions has been reported in
the literature
\cite{Ayon-Beato:ub,Ayon-Beato:ec,Ayon-Beato:1999rg,Ayon-Beato:2000zs}.
The research on this subject started with the pioneering work of
Bardeen \cite{Bardeen68} who proposed the first regular black hole
model (see also
Refs.~\cite{Borde:1994ai,Borde:1996df,Ayon-Beato:2000zs}). The
Bardeen model is a regular black hole satisfying the weak energy
condition and it was a crucial guidance on the ulterior
investigations related with spacetime singularities (see
Ref.~\cite[pag.~265]{H-E}). Subsequently, other regular black hole
models were proposed
\cite{Dymnikova:ux,Ayon-Beato:1993,Borde:1994ai,Borde:1996df,%
Barrabes:1995nk,Mars:1996,Cabo:1997rm,Magli:1997mw}. These models
independently of being interesting by themselves are not exact
solutions of the Einstein equations, and consequently, no
recognizable physical sources can be associated with them. This
situation was changed with the discovery of the results of
Refs.~\cite{Ayon-Beato:ub,Ayon-Beato:ec,Ayon-Beato:1999rg,%
Ayon-Beato:2000zs} using nonlinear electrodynamics, where the
first regular \emph{exact} black hole solutions in the literature
were presented.

Some of the further investigations are directly concerned with
these configurations. They include, for example, the study of the
propagation of the photons governed by the related nonlinear
electrodynamics \cite{Novello:2000km}. The phenomenon of vacuum
polarization on these backgrounds has been exhibited for
conformally coupled quantum fields \cite{Matyjasek:2000iy} and in
the case of more general nonminimal couplings \cite{Berej:2002xd}.
Different prescriptions for calculate the energy distributions
have been explored on these spacetimes in order to shed some light
on the energy localization problem \cite{Radinschi:2000gi}.
Another interesting results concerning the role that such
spacetimes can play as explicit realizations of the confinement
mechanism has been obtained by the authors of
Ref.~\cite{Burinskii:2002pz,Burinskii:2002uf}. Additionally, it
has recently proved that such black holes are stable under
external perturbations \cite{Moreno:2002gg}.

In this paper we present a class of four parametric solutions
which, under physically reasonable assumptions, are regular exact
black hole solutions of the Einstein equations coupled to a
nonlinear electrodynamics acting as a source. In the weak field
approximation this class of solutions reduces to the
Reissner--Nordstr\"om one of the usual linear Maxwell theory.
Moreover, the black hole solutions satisfy the weak energy
condition for a particular range of the parameters. Within this
class, for a particular choice of the parameters, one arrives at
our previous reported solution \cite{Ayon-Beato:ub}.

The gravitational field of the derived solution is given by the
metric
\begin{eqnarray}
\bm{g}&=&-\left(1 -\frac{2mr^{\alpha-1}}{(r^2+q^2)^{\alpha/2}}
+\frac{q^2r^{\beta-2}}{(r^2+q^2)^{\beta/2}}\right)
\bm{dt}^2 \nonumber\\
      & &+\left(1 -\frac{2mr^{\alpha-1}}{(r^2+q^2)^{\alpha/2}}
+\frac{q^2r^{\beta-2}}{(r^2+q^2)^{\beta/2}}\right)^{-1} \bm{dr}^2
+r^2\bm{d\Omega}^2, \label{eq:regbh}
\end{eqnarray}
and the related source is the electric field
\begin{equation}
E=q\left(\frac{\alpha{m}\left[5r^2-(\alpha-3)q^2\right]
r^{\alpha-1}}{2(r^2+q^2)^{\alpha/2+2}}
+\frac{\left[4r^4-(7\beta-8)q^2r^2+(\beta-1)(\beta-4)q^4\right]
r^{\beta-2}}{4(r^2+q^2)^{\beta/2+2}}\right), \label{eq:E}
\end{equation}
where $m$, $q$, $\alpha $ and $\beta $ are parameters. In order to
interpret them we exhibit the asymptotic behavior of the solution
up to $O(1/r^5)$, which is given by
\begin{eqnarray*}
-g_{tt}&=&1-\frac{2m}{r}+\frac{q^2}{r^2}+\alpha\frac{mq^2}{r^3}
-\beta\frac{q^4}{2r^4}+O\left(\frac{1}{r^5}\right),\\
E&=&\frac{q}{r^2}+\alpha\frac{5qm}{2r^3}
-\beta\frac{9q^3}{4r^4}+O\left(\frac{1}{r^5}\right).
\end{eqnarray*}
Up to $O(1/r^3)$ we recover the Reissner-Nordstr\"om behavior and,
consequently, the parameters $m$ and $q$ are related with the mass
and the electric charge, respectively. Moreover, from the
asymptotic behavior of the electric field, the parameters $\alpha$
and $\beta$ could be associated with a sort of dipole and
quadrupole moments of the nonlinear source, respectively. In
correspondence with this interpretation, one notices that our
solution reduces, for $\alpha=\beta=0$, to the
Reissner--Nordstr\"om black hole. For $\alpha=3$, $\beta=4$ one
arrives at our previous reported solution \cite{Ayon-Beato:ub},
hence, the present spacetimes can be thought as a generalization
of the regular black holes of \cite{Ayon-Beato:ub}.

Within the null tetrad formalism, with
$\bm{g}=2\bm{e}^1\bm{e}^2+2\bm{e}^3\bm{e}^4$, we define the null
tetrad
\[
\left.
\begin{array}{r}
\bm{e}^1 \\
\bm{e}^2
\end{array}
\right\}=\frac{r}{\sqrt{2}}
\left(\bm{d\theta}\pm{i}\sin\theta\bm{d\varphi}\right),\qquad
\left.
\begin{array}{r}
\bm{e}^3 \\
\bm{e}^4
\end{array}
\right\}=\frac{1}{\sqrt{2}}\left(\frac{\bm{dr}}{\sqrt{-g_{tt}}}\pm
\sqrt{-g_{tt}}\bm{dt}\right).
\]
With respect to this tetrad our solution is algebraically
characterized by three nonvanishing quantities \cite{Chandra:1983}:
the nonzero traceless Ricci tensor components
\begin{eqnarray}
C_{12}=-C_{34}&=&{}-\frac{\alpha{m}q^2
\left[5r^2-(\alpha-3)q^2\right]r^{\alpha-3}}
{2(q^2+r^2)^{\alpha/2+2}}\nonumber\\
              & &{}-\frac{q^2\left[4r^4-(7\beta-8)q^2r^2
+(\beta-1)(\beta-4)q^4\right]r^{\beta-4}}{4(q^2+r^2)^{\beta/2+2}},
\label{eq:C12}
\end{eqnarray}
the nonvanishing Weyl complex coefficient
\begin{eqnarray}
C^{(3)}&=&{}-\frac{m
\left[6r^4-(7\alpha-12)q^2r^2+(\alpha-2)(\alpha-3)q^4\right]
r^{\alpha-3}}{3(q^2+r^2)^{\alpha/2+2}}\nonumber\\
       & &{}+\frac{q^2
\left[12r^4-3(3\beta-8)q^2r^2+(\beta-3)(\beta-4)q^4\right]
r^{\beta-4}}{6(q^2+r^2)^{\beta/2+2}}, \label{eq:C3}
\end{eqnarray}
and the scalar curvature
\begin{equation}
R=-\frac{2\alpha{q}^2m\left[r^2-(\alpha+1)q^2\right]
r^{\alpha-3}}{(q^2+r^2)^{\alpha/2+2}}
+\frac{\beta{q}^4\left[3r^2-(\beta-1)q^2\right]
r^{\beta-4}}{(q^2+r^2)^{\beta/2+2}}. \label{eq:R}
\end{equation}
Since the curvature invariants $R_{\mu\nu}R^{\mu\nu}$ and
$R_{\mu\nu\alpha\beta}R^{\mu\nu\alpha\beta}$ can be evaluated
using these quantities, we conclude from the behavior of the above
expressions at $r=0$, that the solution is regular for the two
following cases: $\alpha\geq3$, $\beta\geq4$, and $|q|=2m$,
$\alpha\geq1$, with $\beta=\alpha+1$.

In this paper we study the case $\alpha\geq3$, $\beta\geq4$. For
these values of the parameters metric (\ref{eq:regbh}) is a black
hole for a certain range of values of the charge. To establish
this last statement we accomplish the substitutions $x=r/|q|$ and
$s=|q|/2m$, and rewrite $g_{tt}$ as
\begin{equation}
-g_{tt}=A(x,s,\alpha,\beta)\equiv1-\frac{x^{\alpha-1}}
{s(x^2+1)^{\alpha/2}}+\frac{x^{\beta-2}}{(x^2+1)^{\beta/2}},
\label{eq:A}
\end{equation}
which, for any nonvanishing value of $s$, with $\alpha$ and
$\beta$ fixed, has a single positive minimum. Thus, for any
$\alpha$ and $\beta$ there exists a single real critical value of
$x$, $x_{\mathrm{c}}$, and one of $s$, $s_{\mathrm{c}}$, to be
determined from $A(x_{\mathrm{c}},s_{\mathrm{c}},\alpha,\beta)=0$
and $\partial_xA(x_{\mathrm{c}},s_{\mathrm{c}},\alpha,\beta)=0$,
namely
\[
u^\beta-u^{\beta-\alpha}\sqrt{u^2-1}/s+u^2-1=0,\qquad
\left[(\alpha-1)u^2-\alpha\right]u^{\beta-\alpha}
-\sqrt{u^2-1}\left[(\beta-2)u^2-\beta\right]s=0,
\]
where $u^2\equiv1+1/x^2$. To solve these equations, one
substitutes $s$ from the first equation into the second one
arriving at $(\alpha-1)u^{\beta+2}-\alpha{u}^\beta
-(\beta-\alpha-1)u^4+[2(\beta-\alpha)-1]u^2-(\beta-\alpha)=0$,
which has only one positive real root for $u$. For instance, the
corresponding critical values for $\alpha=3$, $\beta=4$ are
$s_{\mathrm{c}}\approx0.317$, $x_{\mathrm{c}}\approx1.58$, for
$\alpha=3$, $\beta=5$ are $s_{\mathrm{c}}\approx0.326$,
$x_{\mathrm{c}}\approx1.48 $, for $\alpha=3$, $\beta=6$ are
$s_{\mathrm{c}}\approx0.335$, $x_{\mathrm{c}}\approx1.41$, etc.
For $s<s_{\mathrm{c}}$ the quoted minimum is negative, for
$s=s_{\mathrm{c}}$ the minimum vanishes and for $s>s_{\mathrm{c}}$
the minimum is positive. Since the solution is regular everywhere
the singularities appearing in (\ref{eq:regbh}), for $s\leq
s_{\mathrm{c}}$, due to the vanishing of $A$ are only coordinate
singularities describing the existence of horizons. Consequently,
we are in the presence of black hole solutions for $|q|\leq
2s_{\mathrm{c}}m$. For the corresponding values of the charge we
have, under the strict inequality $|q|<2s_{\mathrm{c}}m$, inner
and event horizons $r_{\pm}$ for the Killing field
$\bm{k}=\bm{\partial/\partial{t}}$, defined by the real solutions
of the equation $-k_\mu{k}^\mu=A=0$. For $|q|=2s_{\mathrm{c}}m$,
the horizons shrink into a single one, corresponding to an extreme
black hole ($\nabla_\nu(k_{\mu}k^\mu)=0$). The extension of the
metric beyond the horizons $r_{\pm}$ becomes apparent by passing
to the standard advanced and retarded Eddington--Finkelstein
coordinates, in terms of which the metric is smooth everywhere for
$\alpha\geq3$, $\beta\geq4$, even in the extreme case.
Summarizing, our spacetime possesses, for $\alpha\geq3$,
$\beta\geq4$, a similar global structure as the
Reissner--Nordstr\"om black hole except that the singularity, at
$r=0$, of this last solution has been smoothed out and $r=0$ is
now simply the origin of the spherical coordinates. As it has been
previously pointed out this fact originates a change in the
spatial topology of spacetime from open to closed
\cite{Borde:1996df}.

For $|q|>2s_{\mathrm{c}}m$, there are no horizons and the
corresponding exact solution for $\alpha\geq3$, $\beta\geq4$
represents a globally regular spacetime in this coordinates.

The fields (\ref{eq:regbh}) and (\ref{eq:E}) are a solution of the
Einstein-nonlinear electrodynamics field equations which follows
form the action \cite{Plebanski:1968,Salazar:ap}
\begin{equation}
S=\int\mathrm{d}^{4}x\sqrt{-g}\left[\frac{1}{16\pi}R
-\frac{1}{4\pi}\left(\frac12P^{\mu\nu}F_{\mu\nu}
-\mathcal{H}(P)\right)\right], \label{eq:action}
\end{equation}
where $R$ is the scalar curvature and $\mathcal{H}(P)$ is the so
called structural function depending on the invariant
$P\equiv\frac14P_{\mu\nu}P^{\mu\nu}$ of the antisymmetric tensor
$P_{\mu\nu}$. In general the structural function also depends on
the other invariant $Q=\frac14{}^{*}\!P_{\mu\nu}P^{\mu\nu}$ where
${}^{*}$ stands for the Hodge dual, but for the static
configurations we analyze here this invariant is zero. In our case
the structural function determining the nonlinear electrodynamics
giving the regular black hole family (\ref{eq:regbh}) is
\begin{equation}
\mathcal{H}(P)=P\frac{\left(1-(\beta-1)\sqrt{-2q^2P}\right)}
{\left(1+\sqrt{-2q^2P}\right)^{\beta/2+1}}
-\frac{\alpha}{2q^2s}\frac{\left(\sqrt{-2q^2P}\right)^{5/2}}
{\left(1+\sqrt{-2q^2P}\right)^{\alpha/2+1}},  \label{eq:H}
\end{equation}
where $s=|q|/2m$ and the invariant $P$ being a negative quantity.
This structural function satisfies the plausible condition of
correspondence to Maxwell theory, i.e., $\mathcal{H}\approx{P}$
for weak fields ($P\ll{1}$).

Variation of action (\ref{eq:action}) with respect to the
antisymmetric field $P_{\mu\nu}$ gives the material or
constitutive equations relating this field to the electromagnetic
one
\begin{equation}\label{eq:P2F}
F_{\mu\nu}=\mathcal{H}_PP_{\mu\nu}.
\end{equation}
The Einstein and nonlinear electrodynamics equations arise from
action (\ref{eq:action}) under variation with respect to the
metric $g_{\mu\nu}$ and the electromagnetic potential $A_\mu$,
respectively
\begin{equation}
G_\mu^{~\nu}=2(\mathcal{H}_PP_{\mu\lambda}P^{\nu\lambda}
-\delta_\mu^{~\nu}(2P\mathcal{H}_P-\mathcal{H})), \label{eq:Ein}
\end{equation}
\begin{equation}
\nabla_{\mu}P^{\alpha\mu}=0.  \label{eq:Max}
\end{equation}

In order to obtain the solution (\ref{eq:regbh}), (\ref{eq:E}), we
consider the following static and spherically symmetric ansatz for
the metric
\begin{equation}
\bm{g}=-\left(1-\frac{2M(r)}r\right)\bm{dt}^2
+\left(1-\frac{2M(r)}r\right)^{-1}\bm{dr}^2+r^2\bm{d\Omega}^2,
\label{eq:spher}
\end{equation}
and the antisymmetric field
$P_{\mu\nu}=2\delta_{[\mu}^t\delta_{\nu]}^rD(r)$. With these
choices the equation (\ref{eq:Max}) integrates as
\begin{equation}
P_{\mu\nu}=2\delta_{[\mu}^t\delta_{\nu]}^r\frac{q}{r^2}
\quad\longrightarrow\quad
P=-\frac{D^2}2=-\frac{q^2}{2r^4},
\label{eq:dielec}
\end{equation}
where we have chosen the integration constant as $q$ since, as it
was previously anticipated, it plays the role of the electric
charge. The evaluation of the electric field
$E=F_{tr}=\mathcal{H}_PD$ from the constitutive equations
(\ref{eq:P2F}) gives just the formula (\ref{eq:E}). The Einstein
equations (\ref{eq:Ein}) corresponding to the $G_t^{~t}$ component
yields
\begin{equation}
M^{\prime}(r)=-r^2\mathcal{H}(P).  \label{eq:tt}
\end{equation}
Substituting $\mathcal{H}$ from (\ref{eq:H}) with $P=-q^2/2r^4$,
and using that $m=\lim_{r\rightarrow \infty }M(r)$ one can write
the integral of (\ref{eq:tt}) as
\begin{equation}
M(r)=m-q^2\int_r^\infty\mathrm{d}y\left(\frac{\alpha{m}y^{\alpha-1}}
{(y^2+q^2)^{\alpha/2+1}}
+\frac{\left[y^2-(\beta-1)q^2\right]y^{\beta-2}}
{2(y^2+q^2)^{\beta/2+1}}\right). \label{eq:int}
\end{equation}
Using the auxiliary variable $u=1+q^2/y^2$, the
$\alpha$--dependent and $\beta$--dependent integrals above can be
expressed as
\begin{equation}\label{eq:Int(u)}
I_\alpha=\frac{\alpha{m}}{2}\int\frac{\mathrm{d}u}{u^{\alpha/2+1}},
\quad\mathrm{and}\quad
I_\beta=\frac{q^2}{2}\int\left[\frac{1}{u^{\beta/2}}
\mathrm{d}\biggl(\frac{1}{y}\biggr)
-\beta/2\frac{\mathrm{d}u}{u^{\beta/2+1}}\frac{1}{y}\right],
\end{equation}
respectively, from which the primitive of both integrals follows
straightforwardly. Thus, one arrives finally at the expression
\begin{equation}
M(r)=\frac{mr^\alpha}{(r^2+q^2)^{\alpha/2}}
-\frac{q^2r^{\beta-1}}{2(r^2+q^2)^{\beta/2}}, \label{eq:M(r)}
\end{equation}
which allows to conclude the regular dependence (\ref{eq:regbh})
for the metric.

A physically plausible requirement is the fulfillment of the weak
energy condition, i.e., the positivity of the local energy density
along any timelike field. If $\bm{X}$ is a timelike field
($X_{\mu}X^\mu=-1$) the local energy density along $\bm{X}$ for
the matter content at the right hand side of Einstein equations
(\ref{eq:Ein}), can be expressed as
\begin{equation}\label{eq:wec}
4{\pi}T_{\mu\nu}X^{\mu}X^\nu=H_{\alpha}H^\alpha\mathcal{H}_P
-\mathcal{H},
\end{equation}
where $H_\alpha\equiv{}^{*}\!P_{\alpha\mu}X^\mu$ is the magnetic
field intensity associated to $\bm{X}$, and it is also a spacelike
field ($H_{\alpha}H^\alpha>0$) since by definition it is
orthogonal to $\bm{X}$. The requirement of positivity of the above
expression allows to conclude that the weak energy condition is
satisfied if $\mathcal{H}<0$ and $\mathcal{H}_P>0$. For the range
of parameters in the structural function (\ref{eq:H}) allowing the
existence of regular black hole solutions the last conditions
imply: $\alpha=3$, $\beta\geq4$. Summarizing, we are in the
presence of regular black hole solutions satisfying the weak
energy condition for $\alpha=3$. We would like to point out that
there is no contradiction with the Penrose singularity theorem
\cite{Penrose:1964wq} for these cases, since, additionally to the
requirement of the null energy condition (positivity of the local
energy density along signals) and the existence of a closed
trapped surface (black hole), such theorem also demands the
existence of a noncompact Cauchy surface, which is absent for our
spacetimes (see the discussion of Ref.~\cite{Borde:1996df}).

Particular attention deserves the previously quoted regular case
$|q|=2m$, $\alpha\geq1$, $\beta=\alpha+1$. This solution is not a
black hole and it satisfies the weak energy condition only for
$\alpha=1$ ($\beta=2$).

It is worthwhile to mention that for $\alpha=\beta=0$ the
considered nonlinear electrodynamics (\ref{eq:H}) contains the
Maxwell theory, and the corresponding solution (\ref{eq:regbh}),
(\ref{eq:E}) reduces to the Reissner--Nordstr\"om black hole.

\begin{acknowledgments}
This work has been partially supported by FONDECYT Grants 1040921,
7040190, and 1051064, CONACyT Grants 38495E and 34222E,
CONICYT/CONACyT Grant 2001-5-02-159 and Fundaci\'on Andes Grant
D-13775. The generous support of Empresas CMPC to the Centro de
Estudios Cient\'{\i}ficos (CECS) is also acknowledged. CECS is a
Millennium Science Institute and is funded in part by grants from
Fundaci\'on Andes and the Tinker Foundation.
\end{acknowledgments}

\end{document}